\documentclass{cernrep}
\begin{document}
\title{Hard Probes in Heavy-Ion Physics}
\author{Thorsten Renk}
\institute{Department of Physics, University of Jyv\"askyl\"a, Jyv\"askyl\"a, Finland}
\maketitle

\begin{abstract}
The aim of ultrarelativistic heavy ion physics is to study collectivity and thermodynamics of Quantum Chromodynamics (QCD) by
creating a transient small volume of matter with extreme density and temperature. There is experimental evidence that most of the
particles created in such a collision form indeed a thermalized system characterized by collective response to pressure gradients.
However, a numerically small subset of high transverse momentum ($P_T$) processes takes place independent of the bulk, with the outgoing
partons subsequently propagating through the bulk medium. Understanding the modification of such 'hard probes' by the bulk medium is an
important part of the efforts to determine the properties of hot and dense QCD matter. In this paper, current developments are reviewed.
\end{abstract}

\section{Introduction}

When one considers the experimental side of ultrarelativistic heavy-ion physics, one has to deal with events with a multiplicity of $O(10.000)$ particles, i.e. quite a complex environment in which to look for interesting physics. This immediately raises the question about the motivation to study such a system.  The fundamental goal of ultrarelativistic heavy ion physics is the study of Quantum Chromodynamics (QCD), the theory of the strong interaction. The QCD Lagrangean is deceptively simple, its essentials can be written out in two lines
\begin{displaymath}
\mathcal{L}_{QCD} = \mathcal{L}_q + \mathcal{L}_\mathcal{G}
= \overline{\Psi}(i\gamma_\mu D^\mu -{\bf m}) \Psi - \frac{1}{4}
\mathcal{G}_{\mu \nu} \mathcal{G}^{\mu\nu}
\end{displaymath}
with 
\begin{displaymath}
\mathcal{G}_{\mu \nu} = (\partial_\mu A_\nu^a - \partial_\nu A_\mu^a
+g f^{abc} A_{\mu,b} A_{\nu,c}) t_a \quad \text{and} \quad D_\mu = \partial_\mu - igt_aA^a_\mu
\end{displaymath}
and yet it gives rise to a plethora of phenomena, among them a non-trivial vacuum with quark and gluon condensates and instanton configurations, the spectrum of hadrons and hadronic resonances, the binding of baryons into nuclear matter and high-energy phenomena such as the appearance of partonic degrees of freedom, jets and the running of the coupling constant $\alpha_s$. In fact, it is quite fair to say that we are currently unable to work out most of the implications of the QCD Lagrangean from first principles. 

One finds two distinct perspectives taken by researchers in order to explore QCD. The first one is in philosophy termed 'reductionism'. It states that the nature of complex phenomena can be reduced to the nature of simpler or more fundamental phenomena. Applied to QCD, this implies selecting situations in which the fundamental degrees of freedom of QCD, quarks and gluons, are most readily apparent, i.e. to a perturbative description of inclusive high $p_T$ scattering processes in which the dynamics of the underlying event can largely be neglected and the running of $\alpha_s$ ensures that a perturbative expansion in terms of weekly interacting quarks and gluons is meaningful. However, this perspective covers only a subset of QCD phenomenology.

In contrast, a different principle is referred to as 'holism'. It states that there are properties of a given system which cannot be determined or explained by the sum of its component parts alone. Instead, the system as a whole determines in an important way how the parts behave. This second principle is relevant if one asks for collectivity in QCD and properties of QCD matter: Parameters like the viscosity of hot QCD matter cannot be in any meaningful way reduced to properties of isolated quarks or gluons. 

Unfortunately, there is no easy way to translate the holistic perspective into a guide for modelling and understanding collectivity in QCD or heavy-ion collisions. At each scale, the relevant degrees of freedom have to be deduced from experiment before the dynamics of the system can be understood. Often, no consensus about even qualitative insight into the relevant phenomena seen in experiment is found. However, if anything, a holistic perspective argues comprehensive modelling, i.e. taking into account all known dynamics of a system, rather than trying to isolate parts. In particular, in order to understand perturbative high $p_T$ processes in the context of heavy-ion collisions, one also needs a good understanding of the bulk dynamics and vice versa.

\subsection{Heavy-ion collisions, QCD thermodynamics and collectivity}

As such, the presence of $O(10.000)$ secondary particles in the final state of a heavy-ion collision does not imply either collectivity or thermalization. These are two somewhat distinct but related question, and the evidence for each of them must be scrutinized carefully. More specifically, collectivity implies that there is substantial final state interaction, i.e. the final state can not be understood as a mere superposition of many independent elementary $pp$-like collisions. On the other hand, thermalization implies that the phase space distribution of observed final state particles is given by their equilibrium expectation.

\subsubsection{The experiments}

The data shown in this paper to illustrate the ongoing efforts to study heavy-ion physics has been obtained to a small part at the CERN SPS in $\sqrt{s} =7.4$ AGeV Pb-Pb collisions in fixed-target experiments and to the larger part in 200 AGeV Au-Au collisions at the Brookhaven National Lab RHIC collider by the STAR and PHENIX collaborations. While SPS data illustrates some of the bulk phenomena, any perturbative phenomena typically require at least the RHIC kinematic reach, although RHIC barely reaches above the onset of such phenomena. The LHC kinematic reach in 5.5 ATeV Pb-Pb collisions is widely expected to enable a much clearer investigation of many of the phenomena described here.

\subsubsection{Evidence for thermalization}

If hadrons are produced in thermal equilibrium in a sufficiently large system, the density $n_i$ of a hadron species $i$ is be calculable as
\begin{equation}
n_i = \frac{g_i}{2 \pi^2} \int_0^\infty \frac{p^2 dp}{\exp [(E_i - \mu_i)/T   ]\pm 1}
\end{equation}
where $g_i$ is the degeneracy factor, $E_i = \sqrt{m_i^2+p^2}$, $\mu_i$ the sum of baryochemical and strange-chemical potential for the hadron $i$ and $T$ the temperature of the system.

For hadronic resonances, the expression should be integrated over the resonance width, and if the resonance is not detected directly in the experiment, the decay products need to be counted with a weight given by the strength of the particular decay channel. In this way, an expression for the relative yield of different hadron species produced in the decay of a thermal system can be found which only depends on the temperature and baryochemical potential $(T, \mu_B)$ (the strange-chemical potential is fixed by the requirement of overall strangeness neutrality).

\begin{figure}[ht]
\begin{center}
\includegraphics[width=10cm]{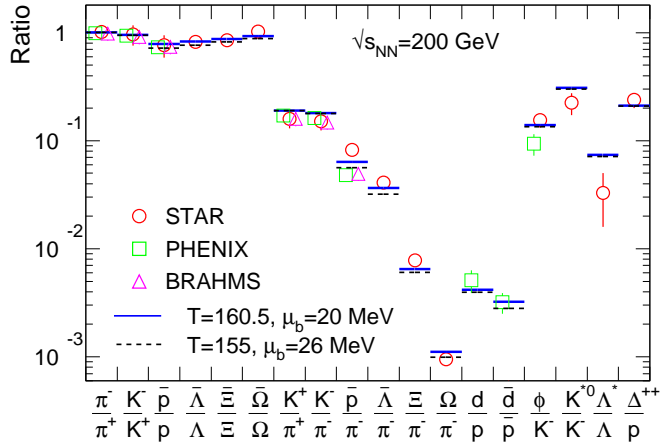}
\caption{Yield rations of different hadron species measured in 200 AGeV Au-Au collisions and compared to equilibrium expectations for two different sets of $(T,\mu_b)$ \cite{StatisticalModel}}
\label{F-Yields}
\end{center}
\end{figure}
 
In Fig.~\ref{F-Yields} the result of two such statistical hadronization models are compared with yield ratios measured in 200 AGeV Au-Au collisions. It is apparent that the equilibrium assumption works remarkably well in describing the data.

One may wonder if this is really evidence for thermalization rather than mere phase space dominance, i.e. the simple fact that if the phase space is large, it gets randomly populated in every event, and averaging over many events then results in apparent thermal-like distribution although in each single event no equilibrium condition holds. In particular, hadron yields from $e^+-e^-$ collisions where there is no reason to suspect equilibration can to some degree also be described by statistical models. However, there is a crucial difference: Statistical models for $e^+-e^-$ collisions require a 'strangeness suppression factor' to take into account the fact that strangeness production is mass suppressed in string-breaking and similar hadronization models. In contrast, the statistical model for Au-Au collisions requires no such factor and still get multistrange ratios like $\Omega/\pi$ correct, which is consistent with the implicit model assumption that the thermal excitation of strangeness is (almost) as strong as the excitation pf $u$ and $d$ quarks. This means that the strangeness production mechanism in heavy-ion collisions is very different from elementary collisions. Moreover, the fact that the Grand Canonical ensemble can be used to compute yield ratios indicates that strangeness is not conserved only locally (i.e. at the production point of the $s\overline{s}$ pair) but that it can propagate over large distances, which may be interpreted as a signal for confinement.

\subsubsection{Evidence for collectivity}

Evidence for collectiviy was discovered already at relatively low energy Pb-Pb collisions at the SPS \cite{NA44}. It was observed that while the transverse mass $m_T = \sqrt{m_{\pi,K,p}^2+P_T^2}$ (with $P_T$ the transverse momentum) spectra of pions, kaons and protons looked each exponential (as characteristic for a thermal system), the slope of the spectrum hardened with particle mass and they could be fit well by the empirical formula
\begin{equation}
\frac{1}{m_T}\frac{dN}{dm_T} \sim \exp[-m_T/T^*]\quad \text{where} \quad T^* = T + m_{\pi,K,p} \langle v_T \rangle^2.
\end{equation}
In this expression, the spectral slope $T^*$ has a component $T$, corresponding to random motion of particles associated with the temperature $T$ and a second component $m_{\pi,K,p} \langle v_T \rangle^2$ corresponding to a collective motion of the volume containing all  hadrons with average velocity $\langle v_T \rangle$, which results in a mass-dependent increase of particle kinetic energy. The underlying picture is that matter moves collectively outward with a radial velocity field, the so-called 'radial flow'.

These arguments have since been refined to describe the medium created in heavy ion collisions in terms of a locally thermalized fluid with the Equation of State (EOS) of QCD where the collective motion is driven by pressure gradients in the fluid. This picture is valid as long as the mean free path of individual particles is much smaller than the medium dimensions --- when this is no longer the case the fluid description ceases to be applicable; the medium 'freezes out' into distinct freely-propagating hadrons which are then detected experimentally.

Viscous relativistic fluid dynamical models \cite{VisHyd1,VisHyd2,VisHyd3} currently represent the state of the art of bulk matter description. They are based on the conservation of the energy-momentum tensor $T^{\mu\nu}$ and any conserved current (usually baryon number only) $j_i^\mu$ in the system,
\begin{equation}
\partial_\mu T^{\mu \nu}  = 0 \qquad \partial_\mu j_i^\mu = 0.
\end{equation}
For a system close to equilibrium, the energy momentum tensor takes the form
\begin{equation}
T^{\mu \nu}  = (\epsilon + p) u^\mu u^\nu - p g^{\mu \nu} + \Pi^{\mu \nu}
\end{equation}
where the first terms represent the structure in the limit of vanishing mean free path of particles in the medium and $\Pi^{\mu \nu}$ represents viscous corrections which come into play when the mean free path is non-vanishing but still small. $\Pi^{\mu \nu}$ has contributions from bulk viscosity $\Pi$ and shear viscosity, where the shear terms dominate the dynamics. These couple to various gradients in the system, i.e.
\begin{equation}
\Pi^{\mu \nu} = \pi^{\mu\nu} + \Delta^{\mu\nu}\Pi \quad \text{and} \quad \pi^{\mu\nu} = \eta \nabla^{\langle \mu}u^{\nu\rangle} - \tau_\pi \left[ \Delta^\mu_\alpha \Delta^\nu_\beta u^\lambda \partial_\lambda \pi^{\alpha \beta} + \frac{4}{3} \pi^{\mu\nu}(\nabla_\alpha u^\alpha)\right]+\dots
\end{equation}
For a stable, causal result, gradients up to 2nd order in the expansion in the (small) deviation from equilibrium need to be included.

A hallmark observable for the fluid-dynamical description is the second harmonic coefficient $v_2$ in the angular distribution of hadrons around the beam axis. It is found as the when the distribution $dN/d\phi$ is expanded in a Fourier series
\begin{equation}
\frac{dN}{d\phi} = \frac{1}{2\pi}[1 + 2 v_1 \cos(\phi) + 2 v_2 \cos(2\phi) + \dots]
\end{equation}
\begin{figure}[ht]
\begin{center}
\includegraphics[width=10cm]{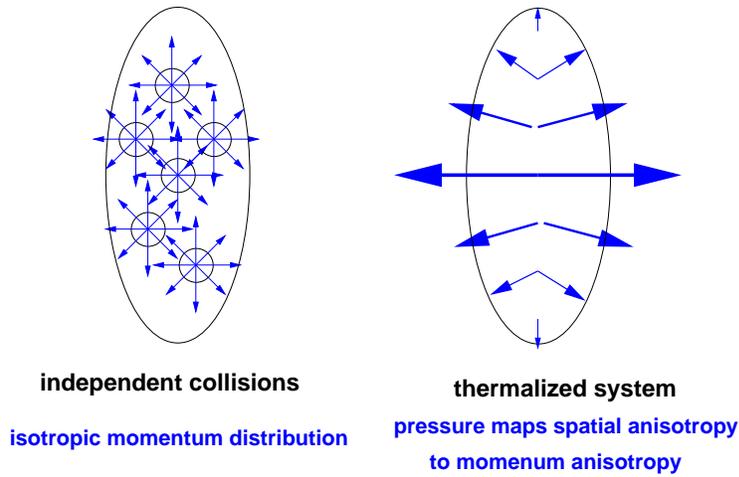}
\caption{Left panel: A superposition of independent collisions does not map a spatial excentricity into a momentum space anisotropy. Right panel: Pressure gradients in a thermalized system can produce such a mapping.}
\label{F-EFlowCartoon}
\end{center}
\end{figure}
The connection of $v_2$ with fluid dynamics is indicated in Fig.~\ref{F-EFlowCartoon}. The figure shows the plane transverse to the beam direction. In non-central collisions, the overlap region of the two colliding nuclei is not circular, but has an almond-like shape, i.e. a spatial excentricity. If there is no collectivity, this excentricity in position space does not imply a corresponding excentricity in momentum space. However, if there is a thermalized medium, spatial excentricity implies stronger pressure gradients in the reaction plane (along the short side) than out of plane (along the long side), and these in turn lead to an excentricity of the particle distribution in momentum space which is manifest as a finite value of $v_2$, the so-called 'elliptic flow'. The crucial test for any fluid dynamical model is therefore its ability to reproduce the experimentally measured value of $v_2$ as a function of both collision impact parameter (or 'centrality') and transverse momentum $P_T$.
\begin{figure}[ht]
\begin{center}
\includegraphics[width=8cm]{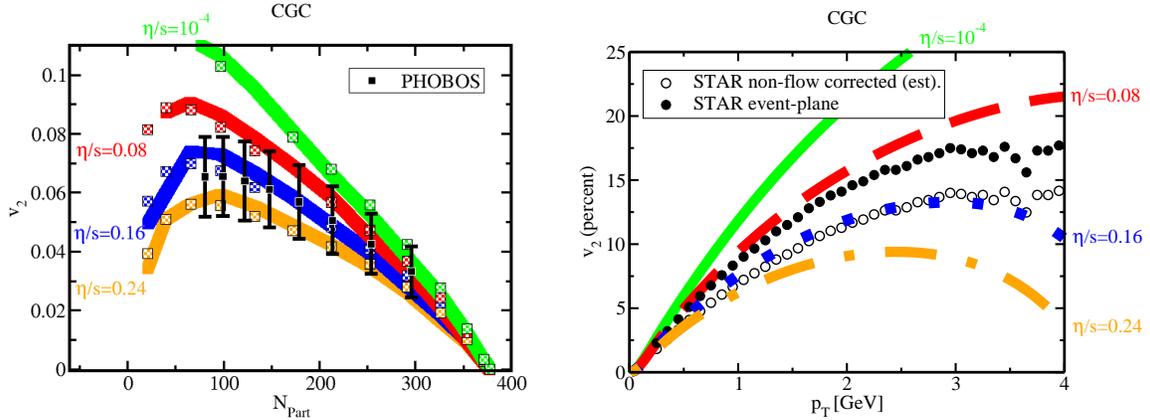}\raisebox{0.3cm}{\includegraphics[width=7.5cm]{v2-talk2}}
\caption{Left panel: The elliptic flow coefficient $v_2$ as a function of the number of collision participants computed for various values of the viscosity over entropy ratio $\eta/s$ and compared with data for 200 AGeV Au-Au collisions. Right panel: $v_2$ as a function of transverse momentum $P_T$ (\cite{VisHyd1}.}
\label{F-EFlow}
\end{center}
\end{figure}
As apparent from Fig.~\ref{F-EFlow}, relativistic viscous hydrodynamics is well able to account for the observed elliptic flow, however only with an almost vanishing viscosity/entropy density $\eta/s$ ratio. This implies that the system is the most ideal fluid observed so far --- superfluid Helium has a ten times larger value for $\eta/s$! Microscopically, the mean free path of particles must therefore be extremely small to reduce viscous corrections to almost zero, which in turn is evidence for a very high degree of collectivity.

\subsection{Bulk matter and probes}

While the observation of yield ratios or elliptic flow constitude evidence that there is collective, thermalized QCD matter, it does not follow that all particles in a heavy-ion collision are part of the thermal system. First, in any collision of two nuclei, there may be individual nucleons not colliding. These so-called 'spectators' continue along the beam direction and are of no further interest. But there may also be secondary particles created in the collision process which are nevertheless not thermal. They fall into two groups:

Due to the smallness of the electromagnetic coupling $\alpha_{em}$ relative to $\alpha_s$ secondary photons and leptons have a mean free path $\sim 100$ times larger than partons or hadrons. In consequence, they almost always escape from the medium without rescattering. In addition, particle production can occur in hard processes with $p_T \gg T$. While the final state partons emerging from such a hard process are strongly interacting, due to a separation of scales their production is calculable as in vacuum, and if $p_T$ is sufficiently large the medium lifetime is insufficient to thermalize them. Experimentally, at RHIC energies above 6 GeV the spectra are no longer described by a thermal distribution but by pQCD. However, in both cases non-termalized particles can serve as 'probes' of the termalized bulk: In the case of electromagnetically inetracting particles, the detailed conditions at their production point in the bulk matter are unknown, but since these particles do not undergo final state interaction, they may be deduced from a measurement. In the case of high $p_T$ processes, the production rate is calculable, but the final state interaction reflects the unknown conditions in the medium. In both cases, important information can be gained.

\begin{figure}[ht]
\begin{center}
\includegraphics[width=8cm]{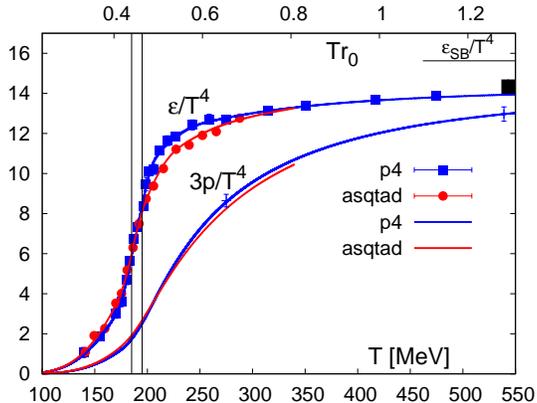}
\caption{The equation of state $p(T)$ of QCD as computed in lattice gauge theory simulations of scaled pressure $p$ and energy density $\epsilon$ as a function of temperature $T$ \cite{EOS}.}
\label{F-EOS}
\end{center}
\end{figure}

We thus arrive at the following picture: The majority of secondary particle production (about 95\%) in a heavy-ion collision ends up as collective, thermalized bulk QCD matter which is well described using almost ideal relativistic fluid dynamics. Its behaviour is largely governed by the EOS of QCD (see Fig.~\ref{F-EOS} for the EOS obtained in lattice gauge theory \cite{EOS}), which exhibits a phase transition around 180-190 MeV temperature from confined matter with hadronic degrees of freedom below the transition and deconfined partonic matter above. Associated with the deconfinement transition is also the restoration of chiral symmetry. Aim of the experiments is to study the detailed properties of this bulk.

Towards this end, it is useful to observe non-thermal particles. These are usually rare, but they have the advantage that either their production or their final state interaction is known, which can be utilized to deduce information about bulk properties which cannot be obtained otherwise. In the following, we will focus on the role of high $p_T$ processes as probes in the context of heavy-ion physics adn do not discuss electromagnetic probes further.

\section{Jet tomography}

'Tomography' is an expression well known from e.g. X-ray imaging. The basic idea is to shine a beam of raditation from a source with known properties through a material of unknown properties and infer from the modification of the radiation after passage through the object its density distribution. The basic idea of jet tomography in the context of heavy-ion collisions is similar. Since a hard process at sufficiently high $p_T$ probes spacetime scales at which the concept of a medium does not even apply, the production rate of hard partons can be calculated perturbatively as in vacuum. The subsequent evolution of the outgoing parton however happens in the medium, such that information about the medium can in principle be deduced from the modification of high $P_T$ hadron yields and correlations.

In practice, there are several complications. First, neither the probe production vertex position nor the probe momentum are known on an event by event basis, but only probabilistically. Second, as seen previously, the medium density distribution is not static, but evolves dynamically. Third, unlike in the case of X-ray tomography, there is no clean separation between medium and probe --- an individual quark or gluon cannot be tagged to be either part of a jet or part of the medium, instead a scale separation in momentum is needed. And finally, since the microscopic dynamics of the medium is not known, the details of parton-medium interaction are unknowns in addition to the evolution of the density distribution.

\subsection{Transport coefficients}

Transport coefficients are a means to parametrize the unknown details of parton-medium interaction. Imagine the passage of many hard partons with their momenta along the $z-$axis through a medium. After the passage through the medium, each momentum will be changed. First, there may be a loss of momentum along the $z-$direction since medium constituents will on average have smaller $p_z$ than a hard parton, so momentum is on average transferred from the hard parton to the medium in each interaction. The mean rate of longitudinal momentum loss per unit length can be called $\hat{e}$, the variance around this mean value $\hat{e_2}$.

In addition, there is also the possibility that a parton picks up a random momentum transverse to its original axis. For symmetry reasons, the mean value of transverse momentum after passage through the medium is zero, but the variance can be cast into the form of a transport coefficient $\hat{q}$.

These transport coefficients are in principle measurable, but they can also be calculated for any given microscopical model of a medium, and if their values are known, they strongly constrain the possible models of the medium. Often, models of parton-medium interaction are formulated in terms of transport coefficients rather than in terms of a microscopical formulation of the medium.

\subsection{Medium-modified fragmentation function and energy loss}

Unfortunately, the objects created in hard QCD processes are almost never high $p_T$ on-shell partons as tacidly assumed above. Usually, highly virtual partons are instead created which radiate gluons or split into $q\overline{q}$ pairs and develop subsequently into a shower of partons, which eventually hadronizes. In pQCD calculations, this is represented by the fragmentation function $D^{vac}_{f\rightarrow h}(z, \mu^2)$ which stands for the yield of hadrons $h$ given a parton $f$ with the hadron taking the momentum fraction $z$ of the parton when the process happens at a momentum scale $\mu$. The fragmentation function contains a non-perturbative part (the hadronization) which cannot be calculated from first principles, but the scale dependence on $\mu$, respresenting the parton shower evolution, can be computed in pQCD.

If one estimates the formation time of virtual partons in the shower, or of final state hadrons, then one finds $\tau \sim E/Q^2; E/m_h^2$ where $Q$ is the parton virtuality and $m_h$ the hadron mass respectively (this expression is derived by realizing that the formation time is the inverse of the energy in a particle's own restframe, and by boosting with the $\gamma$-factor to the lab frame). If one inserts numbers for typical RHIC or LHC kinematics, one finds that while the partonic evolution timescale is similar to medium evolution timescales, the hadronization times are much longer, i.e. the partonic evolution takes place inside the medium and is expected to be modified by it, whereas the hadronization process largely takes place outside of the medium and is hence unmodified. This is on the one hand a promising result, as it gives rise to some hope to calculate the parton-medium interaction perturbatively. On the other hand, it means that the complicated pattern of QCD radiation and its modification by the medium will be a crucial ingredient in the parton-medium interaction model.

It is however possible to simplify the problems. Single inclusive hard hadron production is dominated by events in which most of the momentum of a parton shower flows through a single parton, i.e. there is very little momentum carried by radiated gluons. In this particular case, the fragmentation function largely represents the hadronization of that leading parton, and the medium effect corresponds to a shift in the leading parton energy, i.e. 'energy loss'. If hadronization happens outside the medium, one can assume that the two factorize, and in this particular limit medium induced energy loss becomes a useful concept.

\subsection{From energy loss model to observables}

The basic ingredient to any parton-medium interaction model is thus an expression which describes how the kinematics of a propagating parton and possibly also its gluon radiation changes under the assumption of a particular set of degrees of freedom in the medium. Examples for such models for the elementary interaction including pQCD elastic interactions and induced radiation will be discussed later. 

However, the elementary parton-medium interaction is not observable. Let us discuss the steps going from this expression to an observable quantity for a simple example of single inclusive hadron production. First of all, there may be any number of elementary interactions iterated along a given parton path, and they may be correlated or uncorrelated. Suppose that an elementary radiation process results in the spectrum $\frac{dI(\omega_i)}{d \omega}$ of medium-induced gluon ratiation at energy $\omega$. Assuming that $n$ \emph{independent} interactions along the parton path through the medium can take place, the probability density $P(\Delta E)$ that the total radiated energy is $\Delta E$is given by Poissonization as
\begin{equation}
P(\Delta E)_{path} = \sum_{n=0}^\infty \frac{1}{n!} \left[ \prod_{i=1}^n \int d \omega_i  
\frac{dI(\omega_i)}{d \omega}\right]
\delta\left( \Delta E - \sum_{i=1}^n \omega_i\right) \exp\left[-\int d\omega\frac{dI}{d\omega}. 
\right]
\end{equation}
However, strictly speaking carrying out the sum over $n$ to infinity cannot be correct in a medium of finite particle content, length and lifetime. Unfortunately, there is no analytical expression taking into account the energy degradation of a parton, finite length corrections and correlations among the reactions --- MC codes or numerical solutions of rate equations have to be utilized.

Even $P(\Delta E)$ for a given parton path does not correspond to an observable quantity since the parton path cannot be known. Thus, one needs to average over all unobserved geometry. Hard vertices for impact parameter {\bf b} have a probability distribution to lie in the transverse plane at $(x_0,y_0)$ which is given by

\begin{equation}
P(x_0,y_0) = \frac{T_{A}({\bf r_0 + b/2}) T_A(\bf r_0 - b/2)}{T_{AA}({\bf b})},
\end{equation}

where $T_A({\bf r})=\int dz \rho_A({\bf r},z)$ and $\rho_A({\bf b},z)$ is the nuclear density distribution characteristic for the nucleus. If the probability of energy loss along a given path (determined by medium, vertex ${\bf r_0} = (x_0,y_0)$, rapidity $y$ and transverse angle $\phi$ is $P(\Delta E)_{path}$ one can define the geometry-averaged energy loss probability distribution as

\begin{equation}
\langle P(\Delta E)\rangle_{T_{AA}} \negthickspace = \negthickspace \frac{1}{2\pi} \int_0^{2\pi}  
\negthickspace \negthickspace \negthickspace d\phi 
\int_{-\infty}^{\infty} \negthickspace \negthickspace \negthickspace \negthickspace dx_0 
\int_{-\infty}^{\infty} \negthickspace \negthickspace \negthickspace \negthickspace dy_0 P(x_0,y_0)  
P(\Delta E)_{path}.
\end{equation}

In order to get a hadron spectrum, this expression for the energy loss probability needs to be convoluted with the pQCD parton production spectrum and the vacuum hadronization. Schematically for LO pQCD:

\begin{equation}
d\sigma_{med}^{AA\rightarrow h+X} = \sum_{ijk} {f_{i/A}(x_1,Q^2)} \otimes {f_{j/A}(x_2, Q^2)} \otimes {\hat{\sigma}_{ij 
\rightarrow f+k}} \otimes {\langle P_f(\Delta E) \rangle_{T_{AA}}} \otimes
{D_{f \rightarrow h}^{vac}(z, \mu_F^2)}
\end{equation}

In this expression, $f_{i/A}(x,Q^2)$ stands for the distribution function of parton $i$ in the nucleus at fractional momentum $x$ at scale $Q^2$ and $\hat{\sigma}_{ij \rightarrow f+k}$ for the hard pQCD subprocess  $ij \rightarrow f+k$. The single inclusive hard hadron spectrum $d\sigma_{med}^{AA\rightarrow h+X}$ finally corresponds to an observable quantity.

The steps outlined above are rather generic for any hard probe calculation. Based on a model for the medium degrees of freedom to be tested against the data, an expression for the elementary reaction needs to be derived. With a model for the correlation of such elementary reactions along the parton path, the energy loss probability density, or in a more complete model the medium-modified parton shower can be computed. When a model for the bulk geometry and density evolution is added, the spacetime averaging of the energy loss probability or the medium-modified parton shower can be performed. At this point, tomographical information enters the calculation. Finally, the resulting expression needs to be convoluted with the pQCD parton production expression and a suitable hadronization model to compute an observable. The unavoidable model dependence at each step leads to a sizeable systematic uncertainty and resulting ambiguities in the interpretation of the output, which need to be understood and resolved carefully before any tomographical conclusions are drawn.

\subsection{Parton-medium interaction models}

Let us in the following discuss a few commonly used models for the elementary parton medium interaction.

\subsubsection{Elastic pQCD interactions}

If one models the medium by an approximately free gas of thermal partons, to leading order the interactions of a hard parton with the medium in pQCD become elastic $2 \rightarrow 2$ scattering processes where one incoming parton is hard while the other is thermal. For massless partons, the corresponding cross sections exhibit a $t$-channel singularity for small-angle scattering at low momentum transfer. To regularize the cross section, arguments from thermal field theory (TFT) are often invoked according to which any medium parton acquires a thermal mass $\sim gT$ which screens the singularity. Unfortunately, since $gT$ is a parametric expression for the mass, a prefactor $O(1)$ becomes then an additional free parameter.

Note that in a process like $q\overline{q} \rightarrow gg$ the identity of the hard parton is changed after the scattering process. pQCD interactions with the medium do not only change the kinematics of the hard probe, they may also change the flavour. This is very different for heavy quarks as probes --- due to the lack of thermally excited $c\overline{c}$ pairs at temperatures reached in heavy-ion collisions, the flavour of the hard probe is effectively conserved, a fact which can experimentally be exploited in the comparison of light and heavy quark energy loss. In all cases, the energy lost from the hard parton is carried away by the recoil partons.

\begin{figure}[ht]
\begin{center}
\includegraphics[width=8.0cm]{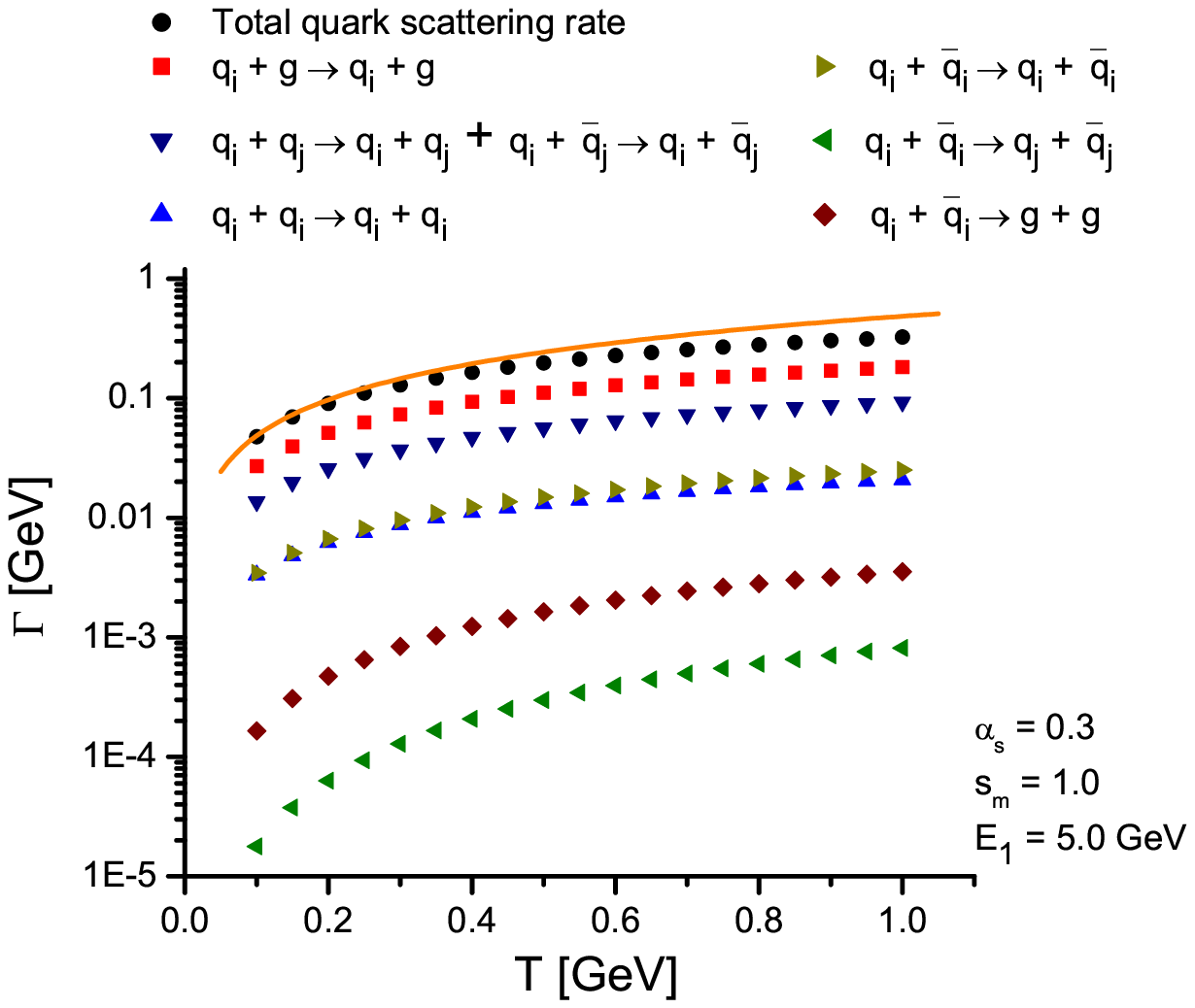}\includegraphics[width=8.0cm]{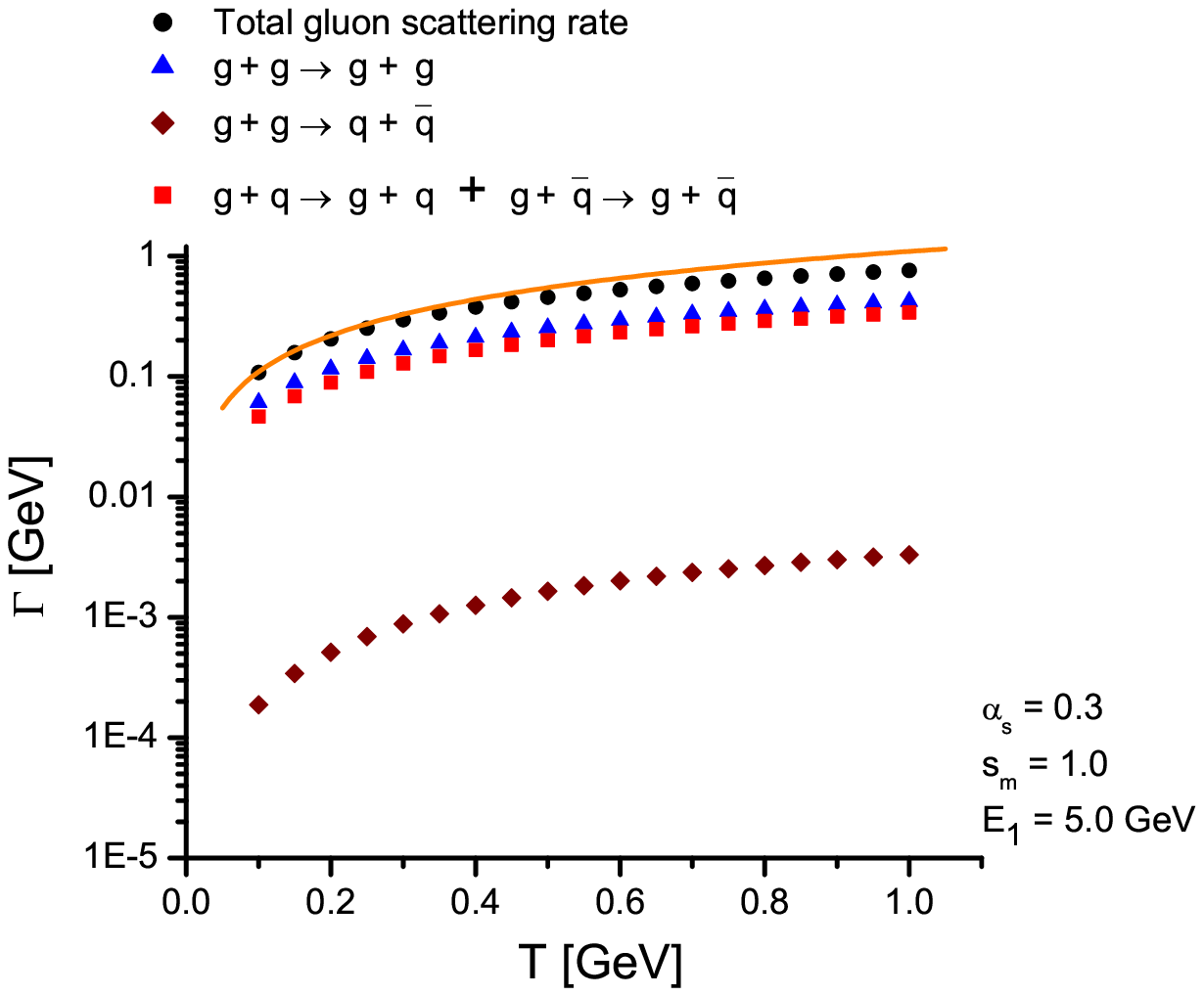}
\caption{Scattering rates for hard quarks (left) and gluons (right) with a medium modelled as a gas of quasi-free thermal quarks and gluons as a function of medium temperature $T$ \cite{ElasticMC}.}
\label{F-elrates}
\end{center}
\end{figure}

The relative importance of the LO pQCD reaction channels primarily depends on the availability of thermal scattering partners. In practice, this means that the gluon channels $qg \rightarrow qg$ and $gg \rightarrow gg$ dominate. The reaction rates $\Gamma$ show some temperature dependence, which is largely determined by the density increase in the medium with temperature. Figure \ref{F-elrates} shows the rates of quark scattering and gluon scattering in the various subchannels as a function of medium temperature as calculated for a MC model of pQCD elastic energy loss \cite{ElasticMC}.

\subsubsection{Medium-induced radiation}

A different mechanism by which a hard parton may lose energy is QCD radiation and branching (in practice this is dominated by soft gluon radiation). The underlying mechanism is that the interaction with the medium opens a kinematic window for radiation which is not available in vacuum. Examples for such a process are an increase in timelike virtuality of the hard parton after a medium interaction followed by a radiation, or a space-like gluon from the virtual cloud surrounding the hard parton color charge being put on-shell by a medium interaction.

pQCD radiation is difficult to compute in general, but the essential scales can be outlined quite easily \cite{QuenchingWeights}: It is crucial to recognize that a radiated gluon cannot decohere instantaneously from the wave function of the parent parton, but that there is a formation time associated with the process. If the energy of the radiated gluon is $\omega$ and the virtuality scale (converted into transverse momentum by the branching process) of the process is $Q$, the formation time can be estimated to be $\tau \sim \omega/Q^2$, and for an object moving with the speed of light this also corresponds to a coherence length $\tau \sim L$.

The virtuality picked up from the medium during the formation time can be estimated using the transport coefficient $\hat{q}$ as $Q^2 \sim \hat{q} L$ and inserting the expression for the coherence length this yields $Q^2 \sim \hat{q} \omega  /Q^2$. This can now be solved for $\omega$ to get the typical radiation energy $\omega_c = \hat{q} L^2$ which grows quadratically with the length.

The energy spectrum of radiated gluons per unit pathlength can then be estimated by noting that one has to sum coherently over all interactions during the coherence time $\tau_{coh}$, thus a factor $\lambda/\tau_{coh}$ (where $\lambda$ is the mean free path) which accounts the average number of scatterings during the coherence time appears in front of the expression for the gluon energy spectrum in a single scattering $ \omega \frac{dI_{1 scatt}}{d\omega dz}$ (i.e. the incoherent limit) which we parametrically take to be $\alpha_s/\lambda$. Thus,
\begin{equation}
\omega \frac{dI}{d\omega dz} \sim \frac{\lambda}{\tau_{coh}} \omega \frac{dI_{1 scatt}}{d\omega dz} \sim \frac{\alpha_s}{\tau_{coh}} \sim \alpha_s 
\sqrt{\frac{\hat{q}}{\omega}}; \quad \text{incl. phase space} \sim \sqrt{\frac{\omega_c}{\omega}}.
\end{equation}
Integrating this expression up to the typical energy scale $\omega_c$ and assuming it dies out above, we indeed find a quadratic dependence of the total mean radiated energy $\langle \Delta E_{rad} \rangle $ on the pathlength as
\begin{equation}
\langle \Delta E_{rad} \rangle = \int_0^\infty d\omega \omega \frac{dI}{d\omega} \sim  \int_0^{\omega_c} d\omega \sqrt{\frac{\omega_c}{\omega}} \sim
\omega_c \sim \frac{\hat{q}}{2}L^2.
\end{equation}
This is crucially different from any incoherent process like the pQCD elastic scattering where we find
\begin{equation}
\langle\Delta  E_{el} \rangle = L\frac{1}{\lambda} \langle \Delta E_{el}\rangle.
\end{equation}
\begin{figure}[ht]
\begin{center}
\includegraphics[width=8cm]{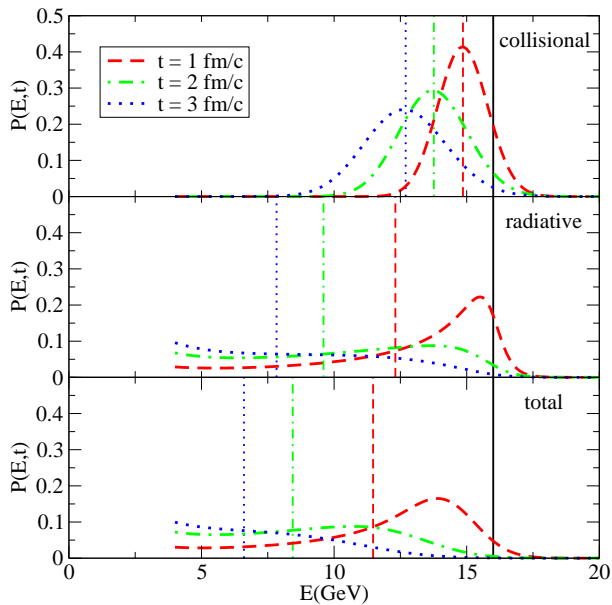}
\caption{Probability of a quark with initial energy $E=16$ GeV embedded in a medium with $T=400$ MeV to have energy $E$ after time $\tau$ considering elastic (collisional) and radiative energy loss as well as their sum computed in the Arnold-Moore-Yaffe (AMY) formalism \cite{AMY}.}
\label{F-AMY}
\end{center}
\end{figure}
A comparison between the effect of elastic and radiative QCD energy loss is shown in Fig.~\ref{F-AMY} where for a medium with 400 MeV temperature the probability of a quark with initial energy $E=16$ GeV to be found with energy $E$ is shown as a function of time. While elastic reactions lead to a downward shift of the mean energy which is constant in time and some moderate broadening around this mean value, radiative energy loss shows a significantly different functional form with a high probability of strong radiation. From the combined figure, one can nicely observe that initially for small pathlengths elastic energy loss is dominant while for large pathlength and late times radiative energy loss becomes more important.

In any realistic situation where pQCD is applicable, there are of course both elastic and radiative processes. In models, it is chiefly the assumed mass of the scattering centers which determines the relative strength. If one assumes almost massless quarks and gluons, the medium is very efficient in taking energy in the recoil of partons and elastic energy loss is a large contribution. Conversely, if one assumes heavy thermal quasiparticles or larger correlated regions of color charge as degrees of freedom then the elastic contribution is much suppressed. The relativ balance of elastic and radiatve energy loss is therefore an important hint with regard to the nature of the microscopic properties of the medium.

\subsubsection{Strong coupling}
 
While arguments can be made that there is a large momentum scale in the pQCD shower evolution of a hard parton in the medium, i.e. the parton virtuality, which allows a perturbative treatment, no such argument can be made for the medium itself or to the coupling of soft, radiated gluons with the medium. While these are often treated perturbatively, it has to be understood that this is an ill-justified ad-hoc assumption.

In contrast, using gauge-gravity duality, the so-called 'AdS/CFT correspondence' \cite{AdS}, it is possible to compute observables in a strongly coupled $N=4$ super-Young-Mills theory. While the particle content of this theory is different from QCD, and QCD exhibits neither conformal invariance nor supersymmetry, there are reasons to believe that the finite temperature sector of both theories is sufficiently similar. 

A strongly coupled medium cannot be described in terms of quasi-particles. Instead, a drag force for propagating quarks appears and the momentum lost from a hard parton excites soundwaves in the medium. In phenomenological models of energy loss \cite{AdS1,AdS2}, this leads to an approximate dependence $\Delta E_{sc} \sim L^3$, yet again different from the coherent radiative and incoherent elastic pQCD scenario. These differences can be exploited in experiment to distinguish the possible scenarios.

\subsection{In-medium shower evolution}

In the leading-parton energy loss approximation, it is sufficient to iterate the elementary radiation process. When also subleading hadrons are considered, the medium interaction needs to be treated on top of a vacuum shower. Typically, this is done in MC codes. In the following, we will illustrate this at the example of the code YaJEM \cite{YaJEM1,YaJEM2} which is based on the PYSHOW \cite{PYSHOW} code for the QCD vacuum shower (other MC codes for medium-modified shower include JEWEL \cite{JEWEL} or Q-PYTHIA \cite{Q-PYTHIA}).

\subsubsection{The unmodified shower}

In the MC picture, the QCD shower is modelled as an iterated series of splittings $a \rightarrow b,c$ where $a$ is a high virtuality parton whereas $b$ and $c$ have lower virtuality $Q$. The evolution variables are the virtuality scale in terms of $t = \ln Q^2/\Lambda_{QCD}^2$ and the fractional momentum $z$ where for the parton energies $E_b = z E_a$ and $E_c = (1-z) E_a$ holds. The differential probability for a branching at scale $t$ is given by the integral of the branching kernel over all kinematically allowed values of $z$ as
\begin{equation}
I_{a\rightarrow bc}(t) = \int_{z_-(t)}^{z_+(t)} dz \frac{\alpha_s}{2\pi} P_{a\rightarrow bc}(z).
\end{equation}
The branching kernels can be computed in pQCD for the different subprocesses as
\begin{equation}
\label{E-kernels}
P_{q\rightarrow qg}(z) = \frac{4}{3} \frac{1+z^2}{1-z} \quad P_{g\rightarrow gg}(z) = 3 \frac{(1-z(1-z))^2}{z(1-z)} \quad P_{g\rightarrow q\overline{q}}(z) = \frac{N_F}{2} (z^2 + (1-z)^2)
\end{equation}
From these expressions, the probability density for the next splitting process of $a$ occuring at a lower scale $t_m$ when coming down from an initial scale $t_{in}$ is given by
\begin{equation}
\label{E-Qsq}
\frac{dP_a}{dt_m} = \left[\sum_{b,c}I_{a\rightarrow bc}(t_m)  \right] \exp\left[ - \int_{t_{in}}^{t_m} dt' \sum_{b,c} I_{a \rightarrow bc}(t') \right].
\end{equation}
i.e. by the probability for the splitting process times the probability that no branching has already taken place before at a higher scale, the so-called 'Sudakov form factor'. Solving these expressions and energy-momentum conservation numerically corresponds to the PYSHOW algorithm.

\subsubsection{Spacetime picture and parton-medium}

While the vacuum shower evolution is computed in momentum space only, an in-medium shower needs the additional information where and when the evolving medium is probed. Based on uncertainty relation arguments for the formation time of a radiated parton, one can estimate the average formation time for a branching $a \rightarrow b,c$ as
\begin{equation}
\langle \tau_b  \rangle= \frac{E_b}{Q_b^2} - \frac{E_b}{Q_a^2}
\end{equation}
which in a MC formulation can be randomly distributed with a probability density
\begin{equation}
P(\tau_b) = \exp\left[- \frac{\tau_b}{\langle \tau_b \rangle}  \right].
\end{equation}
Under the additional assumption that partons move on eikonal trajectories, this provides a spacetime picture of the shower so that the transport coefficients of the medium at any given step of the shower evolution can be obtained.

Currently, there are several models approximating the parton-medium interactions in the shower. Some are based on explicitly changing the parton kinematics. For instance, assuming the medium acts predominantly by increasing parton virtuality as
\begin{equation}
\Delta Q_a^2 = \int_{\tau_a^0}^{\tau_a^0 + \tau_a} d\zeta \hat{q}(\zeta)
\end{equation}
leads to additional medium-induced radiation (referred to as RAD within YaJEM), whereas
\begin{equation}
\Delta E_a = \int_{\tau_a^0}^{\tau_a^0 + \tau_a} d\zeta D \rho(\zeta)
\end{equation}
corresponds to a drag force like interaction (DRAG). Yet a different possibility is to approximate the medium-interaction by a modification of the QCD splitting kernels Eq.~(\ref{E-kernels}) to enhance low $z$ gluon radiation (FMED in YaJEM, this is also used in JEWEL \cite{JEWEL} and Q-PYTHIA \cite{Q-PYTHIA}).
\begin{figure}[ht]
\begin{center}
\includegraphics[width=8cm]{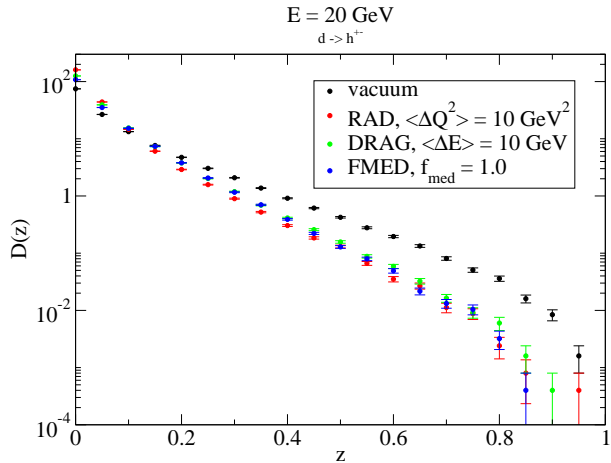}
\caption{Comparison of the MMFF with the vacuum fragmentation function of a 20 GeV $d$-quark into charged hadrons for three different parton-medium interaction models in YaJEM \cite{YaJEM2}.}
\label{F-MMFF}
\end{center}
\end{figure}
All these scenarios lead to a medium-modified fragmentation function (MMFF) which is rather similar as far as the leading shower parton is concerned. As seen in Fig.~\ref{F-MMFF}, there is a marked depletion of large $z$ which corresponds to leading parton energy loss, given that the average $z$ probed in a computation of the hadron spectrum at RHIC kinematic conditions is $z \sim 0.7$. In the radiative models (RAD and FMED) this is compensated by an increase in hadron production at low $z$, whereas no such increase is seen in the DRAG model. This is not unexpected, given the assumption that in this model 'lost' energy from the leading parton excites soundwaves in the bulk medium. Experimentally, one can try to exploit this difference to determine which mechanism for the redistribution of energy lost from the leading parton is realized in nature.

\section{Observables}

In the previous section, we have seen that there is a large number of possibilities how the parton-medium interaction could be realized, dependent on the relevant degrees of freedom in the medium. At the same time, while there is a broad consensus that relativistic fluid dynamics is a valid framework to describe the dynamics of bulk matter, different implementations of the model do not usually agree in the evolution of medium density they predict. 

The problem is therefore twofold: From a collection of experimental observables, one would like to deduce both information on the microscopical dynamics of parton-medium interaction realized in nature as well as constraints for the fluid-dynamical models of the bulk matter density evolution. At the same time, only very few parameters can be controlled or determined experimentally, among them the collision centrality, the collision energy, the particle type detected and the orientation of particles with respect to the reaction plane.

The comparison of models with data tries to make use of the handles discussed previously to distinguish different scenarios --- chiefly the pathlength dependence of leading parton energy loss, but also the nature of energy redistribution by either gluon radiation or shockwaves.

\subsection{Single-inclusive hadron observables}

The simplest possible measurement is the spectrum of single-inclusive high $P_T$ hadron production in heavy-ion collisions. In order to take out the trivial fact that there are hundreds of p-p-like binary nucleon-nucleon collisions in a heavy-ion collision, usually the spectrum is cast into the form of a ratio, the nuclear suppression factor $R_{AA}$

\begin{equation}
R_{AA}(P_T,y) = \frac{d^2N^{AA}/dp_Tdy}{T_{AA}(0) d^2 \sigma^{NN}/dP_Tdy}
\end{equation}

in which the yield of hadrons in A-A collisions is divided by the number of binary collisions times the yield in p-p collisions. Experimentally, $R_{AA}$ in central collision is found to be roughly 0.2, i.e. about 4 of 5 high $P_T$ hadrons appear to be modified by the medium. The functional form of $R_{AA}$ as a function of the hadron kinematic variables $(P_T,y)$ turns out to be trivial, i.e. $R_{AA}$ is largely flat and the normalization 0.2 is the only  parameter which can be extracted from the data.

It can be shown that the $P_T$-dependence of $R_{AA}$ is largely driven by the shape of the pQCD parton spectrum --- even drastic variations of the functional form of the energy loss model lead to only weak changes in the shape of the resulting $R_{AA}$ \cite{gamma-h}. On the other hand, any model requires a connection between the thermodynamical parameters like temperature $T$, entropy density $s$ or energy density $\epsilon$ and the transport coefficients. Usually, a relation like $\hat{q} = const. \cdot s$ is assumed which involves one free parameter. It follows that $R_{AA}$ for central collisions can be described by almost any model provided that the one parameter is adjusted to the normalization seen in the data, and that any non-trivial test of the parton-medium interaction model or the medium density evolution requires a comparison with more differential quantities.

Comparing with $R_{AA}$ at larger centralities (larger impact parameter) then probes a model in a non-trivial way. At more peripheral collisions, the normalization of $R_{AA}$ increases for two different reasons. First, the average density of the medium in terms of available scattering partners reduces, and second the average in-medium pathlength is reduced as the transverse overlap area of the colliding nuclei is decreased. While changes in the average density tend to affect all models of parton-medium interaction in the same way, the models respond to a change in pathlength differently. The pathlength weight is $L$ for elastic/incoherent processes, $L^2$ for coherent radiative processes and $L^3$ for strong coupling. For the models discussed previously one expects the ordering $R_{AA}^{AdS} > R_{AA}^{rad} > R_{AA}^{el}$ for non-central collisions provided all models give an equally good description for central collisions.

An even more constraining observable for the pathlength dependence is to consider $R_{AA}(\phi)$ with $\phi$ the angle of hadrons with the reaction plane in non-central collisions. This allows to vary average pathlength without changing the medium density. For in-plane emission, partons always cross the short side of the almond-shaped overlap region, whereas for out-of-plane emission they cross the long side. However, the difference between in-plane and out-of-plane pathlength also depends on the sharpness of the assumed overlap profile and the speed at which a fluid dynamical evolution expands the surface. Thus, $R_{AA}(\phi)$ always probes a combination of medium model and interaction model.

\begin{figure}[ht]
\begin{center}
\includegraphics[width=8.0cm]{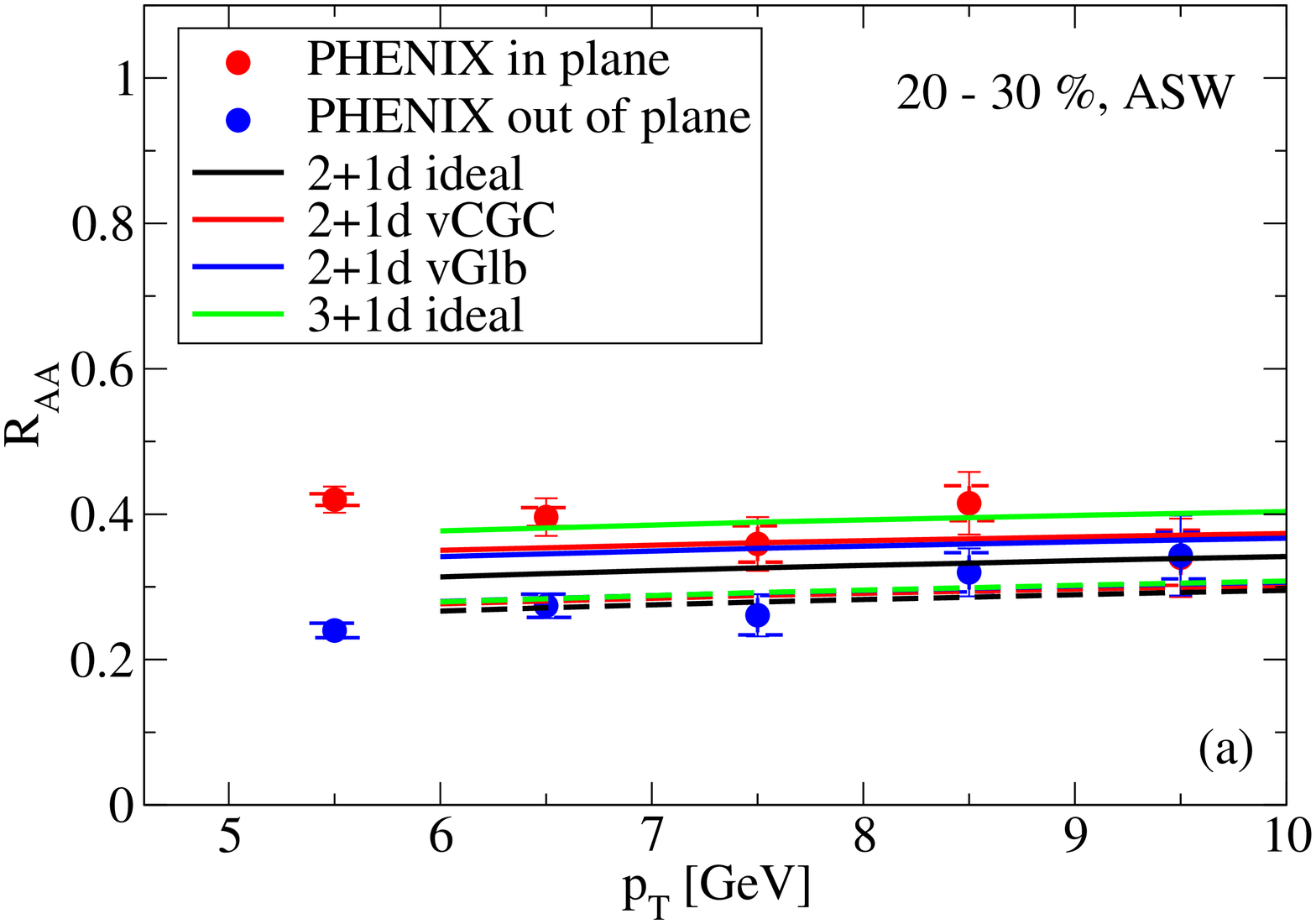}\includegraphics[width=8.0cm]{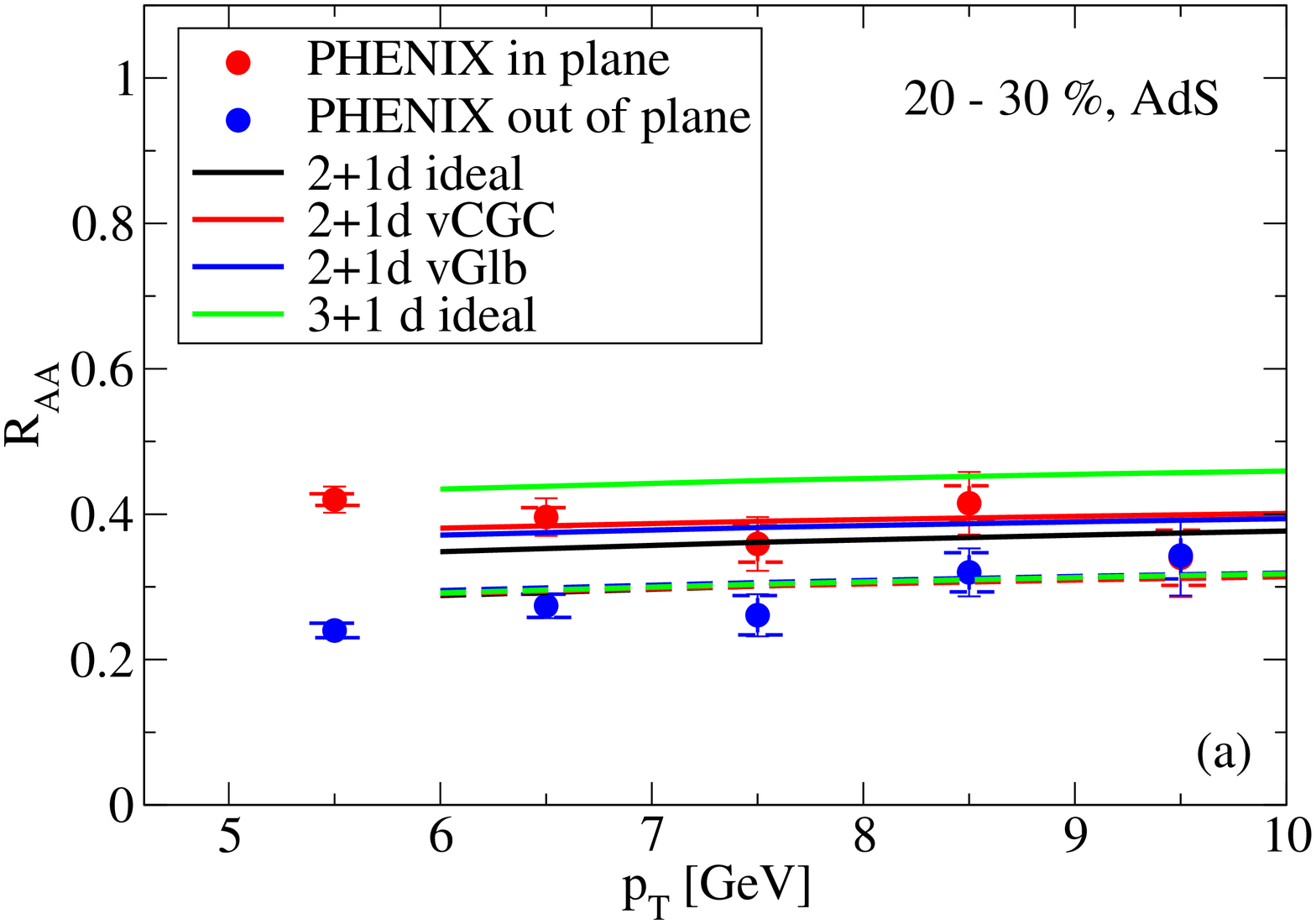}
\caption{Comparison of the nuclear suppression factor calculated for different hydrodynamical models of the medium  for 20-30\% central 200AGeV Au-Au collisions \cite{JetHydroSys} with data \cite{PHENIX-RAA}. Left panel: assuming pQCD radiative energy loss (ASW). Right panel: assuming a strongly coupled medium (AdS).}
\label{F-RAA}
\end{center}
\end{figure}

A systematic investigation of different models for both medium evolution and parton-medium interaction has been made in \cite{JetHydroSys}. Fig.~\ref{F-RAA} shows some of the results compared with PHENIX data \cite{PHENIX-RAA}.

The comparison involves four different models for the medium evolution --- two models without viscosity corrections (ideal hydrodynamics), two with viscosity corrections. The 2+1d ideal hydrodynamical model differs from the 3+1d model mainly by starting and ending the fluid phase earlier. The main difference between the viscous hydrodynamical models is the sharpness of the initial density profile --- while vCGC corresponds to a rather steep density gradient and a well-defined overlap region, vGlb is considerably smoother. All the medium model are constrained by a number of bulk observables.

The main result of the figure is that $R_{AA}(\phi)$ responds in a characteristic way to both the parton-medium interaction model and the medium evolution model and constrains combinations of them. For instance, while the 3+1d ideal fluid dynamics works well with pQCD radiation, the spread between in-plane and out of plane emission is too wide if strong coupling dynamics is assumed. One of the most striking findings (not shown) is that elastic incoherent processes always fail to describe the data, no matter what medium model is assumed. This practically rules out quasi-free quarks and gluons as relevant degrees of freedom in the medium.

As far as the evolution dynamics is concerned, a few trends seem to emerge. For instance, viscous corrections are likely to be important, an early decoupling is not favoured and the initial geometry is surprisingly unimportant. However, given the ambiguity seem from the plots, the need for distinct and more differential observables is clearly apparent.

\subsection{Dihadron and $\gamma$-hadron correlations}

A different way of probing the medium makes use of the fact that most hard processes result in a hard back-to-back parton pair. Coincidence measurements try to recapture this structure in the final state hadron distribution. In particular, usually one triggers on one hard hadrons and then computes the correlation strength of other hadrons in a certain momentum window as a function of angle with the trigger. The typical high $P_T$ correlation function exhibits strength at angles $0$ and $\pi$ --- the first coming from subleading hadrons in the shower which produced the trigger, the second from the recoiling second hard parton shower. 

\begin{figure}[ht]
\begin{center}
\includegraphics[width=10.0cm]{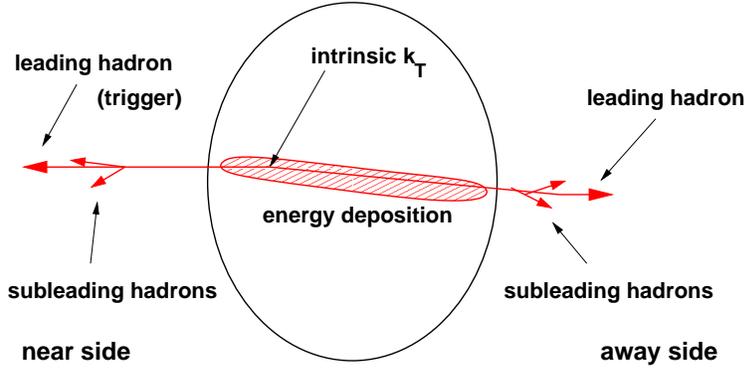}
\caption{Terminology used in back-to-back correlation measurements.}
\label{F-b2b}
\end{center}
\end{figure}

Some terminology and the connection with energy loss is shown in Fig.\ref{F-b2b}. In particular, due to the medium interaction, even a small loss of leading parton energy implies a large suppression due to the steeply falloff of the parton spectrum with $p_T$. Thus, hard trigger hadrons tend to come from regions where they cross as little medium as possible --- this is called 'surface bias'. But since the away side hadron has to come from the same vertex, the implication is that its in-medium pathlength is almost maximized. Thus, one expects a particularly strong sensitivity to the pathlength scaling.

\subsubsection{Hadron-hadron correlations}

\begin{figure}[ht]
\begin{center}
\includegraphics[width=12.0cm]{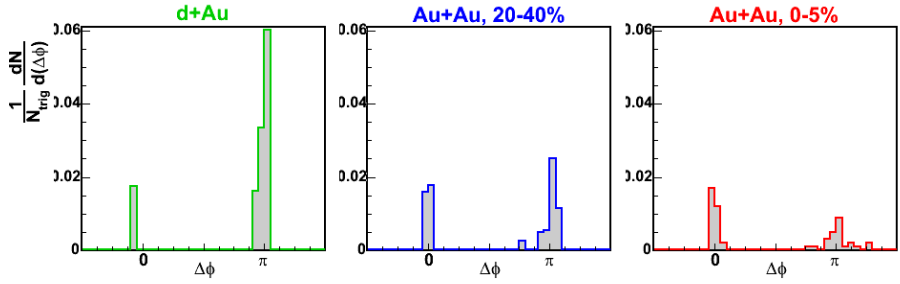}
\caption{Hadronic back-to-back correlation functions as measured by the STAR collaboration \cite{STAR-dijets} for d-Au and two different centralities for Au-Au collisions. The disappearance of the away side correlation for increasing centrality is clearly visible.}
\label{F-b2b1}
\end{center}
\end{figure}

The relevant experimental quantities for correlation measurements are the yield per trigger in a given away side momentum window, and, dervied from that, the suppression factor $I_{AA}$ which is the ratio of the per trigger yield in A-A collisions divided by the per trigger yield in p-p collisions. Experimentally, $I_{AA}$ is found to be on the order of $R_{AA}$, i.e. around 0.2-0.3. Since the trigger yield itself is suppressed (as given by $R_{AA}$, the total suppression of back-to-back events in central 200 AGeV Au-Au collisions is of the order of 95\%! This is quite a dramatic effect and gives rise to 'monojet' phenomena where only one hard parton is observed whereas the energy of the away side parton is completely absorbed and redistributed in the medium (see Fig.~\ref{F-b2b1}).

Back-to-back correlations of high $P_T$ hadrons are typically modelled with MC codes where the resulting back-to-back events are subjected to the set of experimental cuts. Detailed investigations are numerically rather involved. The emerging picture is that h-h-correlations do not seem to add substantial information beyond what can already be gained from $R_{AA}(\phi)$ --- models which describe the single inclusive hadron suppression well also tend to describe the observed dihadron suppression \cite{Dihadron1,Dihadron2}. Similarly, elastic (incoherent) models which fail to reproduce $R_{AA}(\phi)$ fail even more prominently with $I_{AA}$ \cite{Dihadron3}.

\subsubsection{$\gamma$-hadron correlations}

$\gamma$-hadron correlation measurements are suppressed by a factor $\alpha_{em}/\alpha_s$ as compared to h-h correlations and are experimentally harder to do, but they offer one significant advantage: Unlike a trigger hadron which is part of a shower, a trigger photon ideally carries the full information about the kinematics of the event. Thus, one knows what energy to expect on the away side and can determine how much of this energy is recovered in a given angular and momentum window, i.e. one potentially measures the full fragmentation fragmentation function rather than the high $z$ part only. In other words, $\gamma$-h correlations allow to study the mechanism of energy redistribution after that energy is lost from the leading parton and modifications of the fragmentation function.

In practice, there are some complications: Not all hard photons are formed in the primary process. Photons may also be produced as part of a parton shower ('fragmentation photons') and in elastic reactions with the medium such as $q\overline q \rightarrow g\gamma$ ('conversion photons'). However, these effects can be accounted for systematically.

\begin{figure}[ht]
\begin{center}
\includegraphics[width=8.0cm]{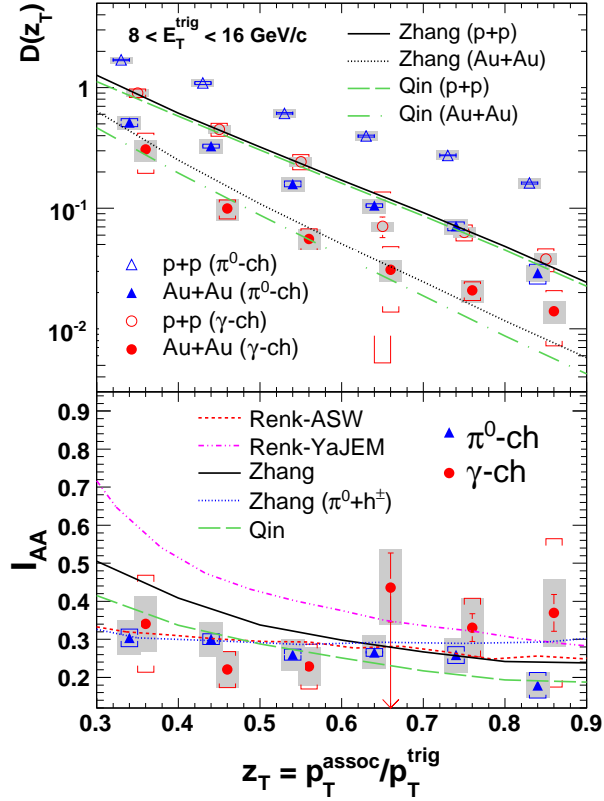}
\caption{$\gamma$-h back-to-back correlation functions as measured by the STAR collaboration \cite{STAR-dijets} in terms of the away side fragmentation function and away side $I_{AA}$ compared with several models.}
\label{F-gamma-h}
\end{center}
\end{figure}

An example of data compared with model calculations is shown in Fig.~\ref{F-gamma-h}. Most models shown are leading parton energy loss models and describe the data well. However, these models make the implicit assumption that energy lost from the leading parton is \emph{not} carried by subleading hadrons in the shower where the experimental procedure would detect it but shifted to very low momenta and large angles. In contrast, YaJEM (as discussed earlier) assumes that energy lost from a leading parton leads to an enhancement low low $z$ gluons in the shower, which after hadronization turns into increased multiplicity at low $z$. This leads to the sharp upward bending of $I_{AA}$ in YaJEM (which would eventually rise above unity for even lower $z$) which is not seen in the data. The absence of any such enhancement is suggestive of non-perturbative energy loss mechanisms which couple directly to bulk matter, e.g. the excitation of soundwaves.

\subsection{Jets}

Jets are a more natural framework to discuss pQCD at high $P_T$ than single inclusive hadron spectra and correlations, as they can be in the absence of a medium defined without strong dependence on complications like hadronization models. Ideally, a jet definition is independent if the jet is treated at the partonic (pQCD) level, at the hadron level or at the detector level. However, in the presence of a medium, jets are much more complicated.

First, the presence of sizeable fluctuating background given by the medium, both in particle number and in energy density, makes jet-reconstruction rather complicated as compared to the p-p case \cite{JetReco}. More specifically, the problem is not so much finding a jet, but assigning the correct energy. But there are also more fundamental conceptual problems. In the absence of a medium, a jet has a representation at the hadron level because the hadrons which carry the original hard parton momentum are created by pQCD branching processes. This is not so in the medium, as part of the energy and momentum can be carried by medium degrees of freedom due to parton-medium interactions. Thus, the original parton is only represented by the flow of energy and momentum in the final state, not by any specific group of hadrons. This may be problematic for sequential recombination algorithms.

Thus, a low $P_T$ hadron may be correlated with a jet for a number of reasons. First, it may be part of the hadronizing parton shower. Second, it may be born in the medium, but have interacted with the medium. And third, it may simply accidentially share a common bias. For instance, unmodified jets tend to emerge perpendicular to the medium surface (because this minimized their in-medium path). Yet at the same time, this is the direction of radial flow which causes other phenomena.

In addition, in computing medium-modified jets in the parton shower language, there is a tacid assumption that hadronization takes place outside the medium. While this is true for light hadrons and for leading shower hadrons, it is certainly not true for heavy or subleading hadrons which would hadronize in the medium. Since we lack a detailed understanding of hadronization even in the vacuum, we cannot compute this part of the jet modification reliably. A simple strategy (with its own pitfalls) is therefore to apply a $P_T$ cut and define jets only above this cut.

While it is expected that studying the medium modification of jets will become a major part of the LHC heavy ion program, currently the field is in its infancy. The following selected results from YaJEM \cite{YaJEM-Jets} should therefore be regarded as a proof of principle only.

\begin{figure}[ht]
\begin{center}
\includegraphics[width=8.0cm]{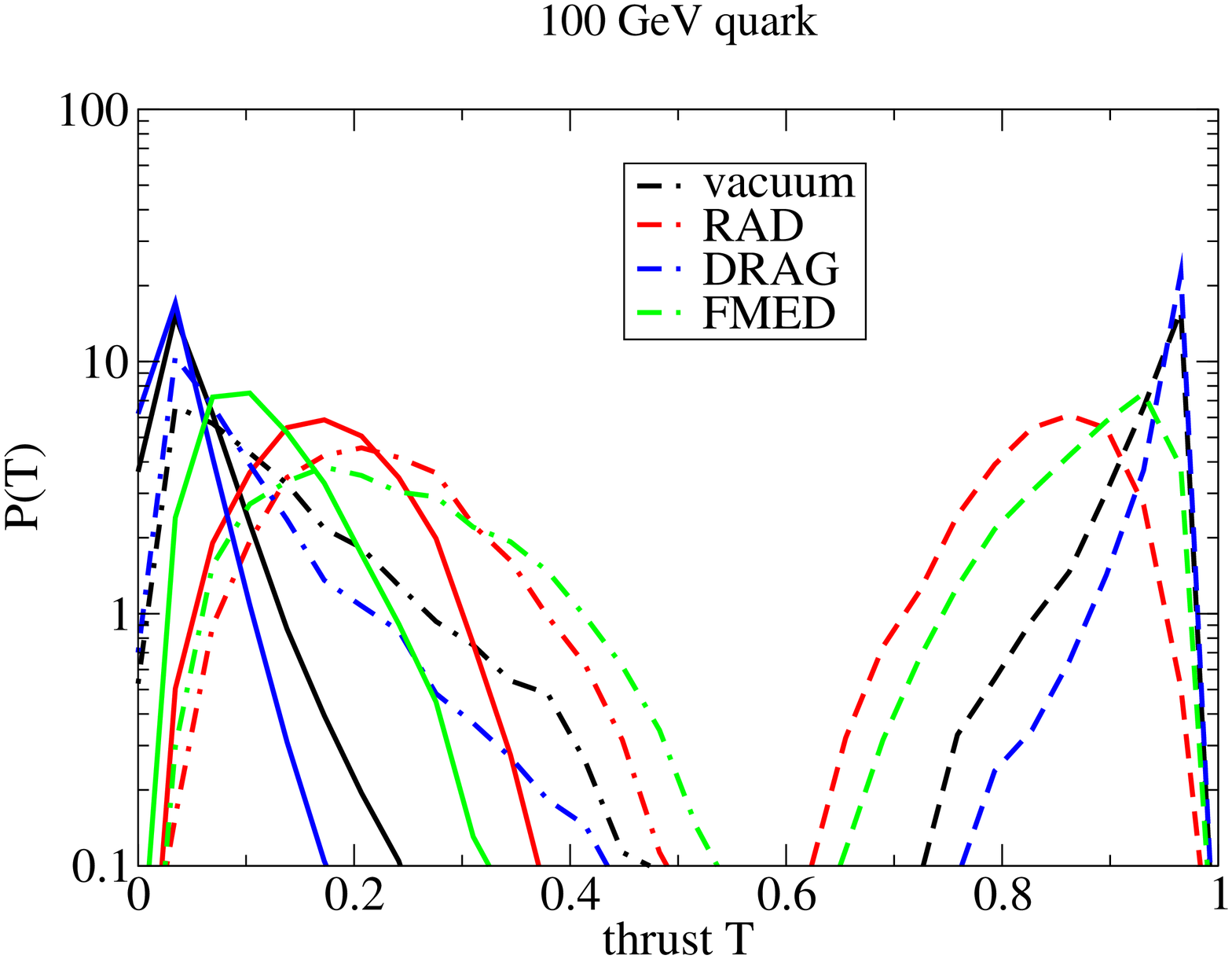}\includegraphics[width=8.0cm]{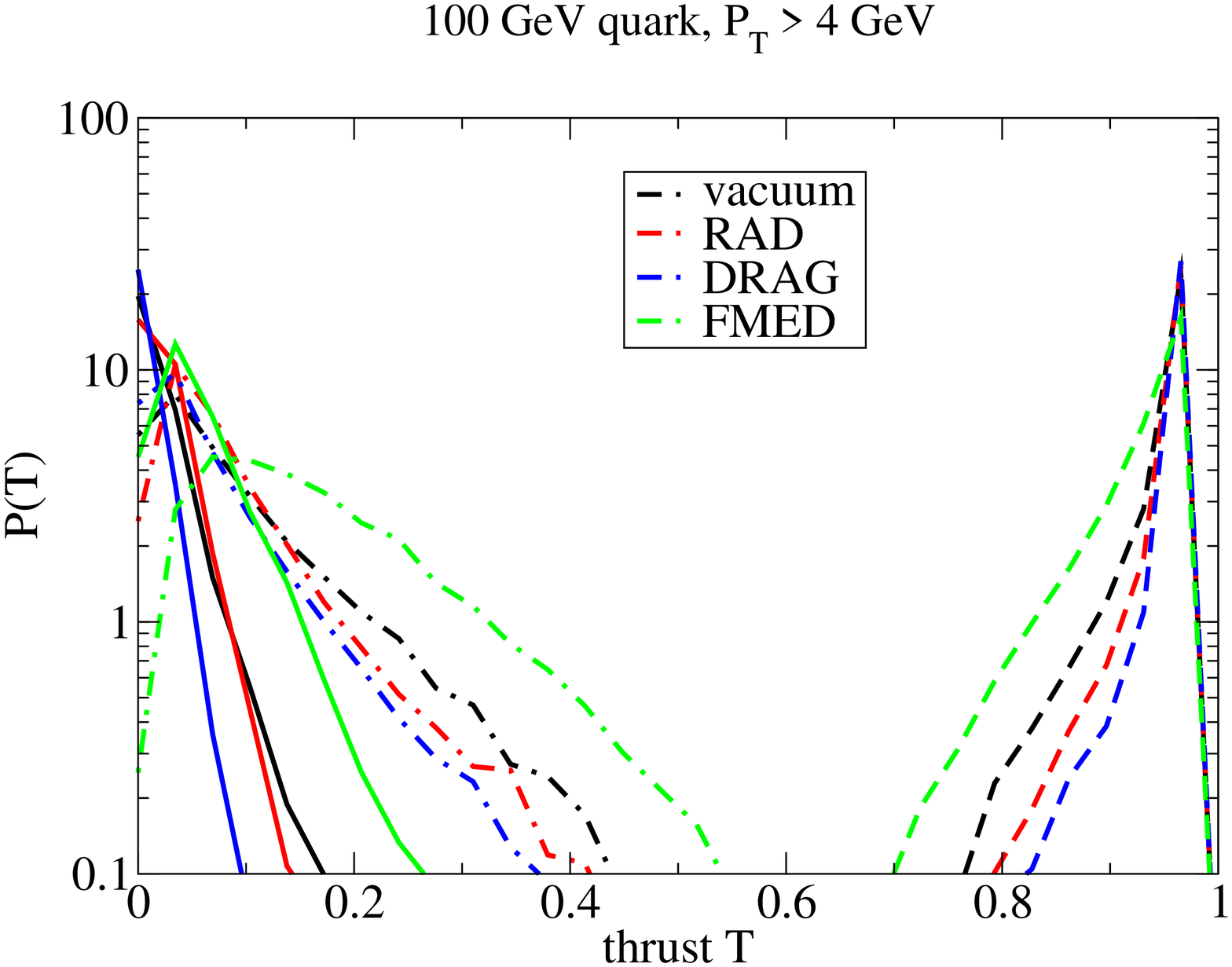}
\caption{Thust distribution of jets from a 100 GeV quark both in vacuum and medium modified as computed within YaJEM \cite{YaJEM-Jets}.}
\label{F-thrust}
\end{center}
\end{figure}

The distribution of thrust, thrust major and thrust minor characterizes jet events in a global way. The distributions are defined as sums over all particles in the event follows:
\begin{equation}
T = \text{max}_{{\bf n}_T} \frac{\sum_i | {\bf p}_i \cdot {\bf n}_T|}{\sum_i|{\bf p}_i|} \quad
T_{maj} = \text{max}_{{\bf n}_T \cdot {\bf n}=0} \frac{\sum_i | {\bf p}_i \cdot {\bf n}|}{\sum_i|{\bf p}_i|} \quad
T_{min} = \frac{\sum_i | {\bf p}_i \cdot {\bf n}_{mi}|}{\sum_i |{\bf p}_i|}
\end{equation}
In particular, thrust is a measure for how spherical an event is. A value of one indicates a pure back-to-back event, whereas a value of 0.5 indicates a completely spherically symmetric event. Fig.~\ref{F-thrust} shows how the presence of a medium as expected for LHC modifies the events within YaJEM. It is clearly seen that the large amount of induced gluon radiation tends to make the event more spherical at low $P_T$ whereas the effect is significantly diluted above a $P_T$ cut of 4 GeV to eliminate medium hadrons from the jet.
An observable which more locally traces the pQCD splitting in the shower evolution is the $n$-jet fraction. This is based on clustering an event with a  sequential recombination algorithm with a set resolution scale $y_{min}$ based on the distance measure
\begin{equation}
y_{ij} = 2 \text{min}(E_i^2,E_j^2) (1-\cos(\theta_{ij})/E_{\text{cm}}^2
\end{equation}
and counting the number of recovered subjets. For a large resolution scale and a back-to-back event, usually 2 jets will be found, but with increasingly fine scale subjets created by early branchings are recovered.
\begin{figure}[ht]
\begin{center}
\includegraphics[width=8.0cm]{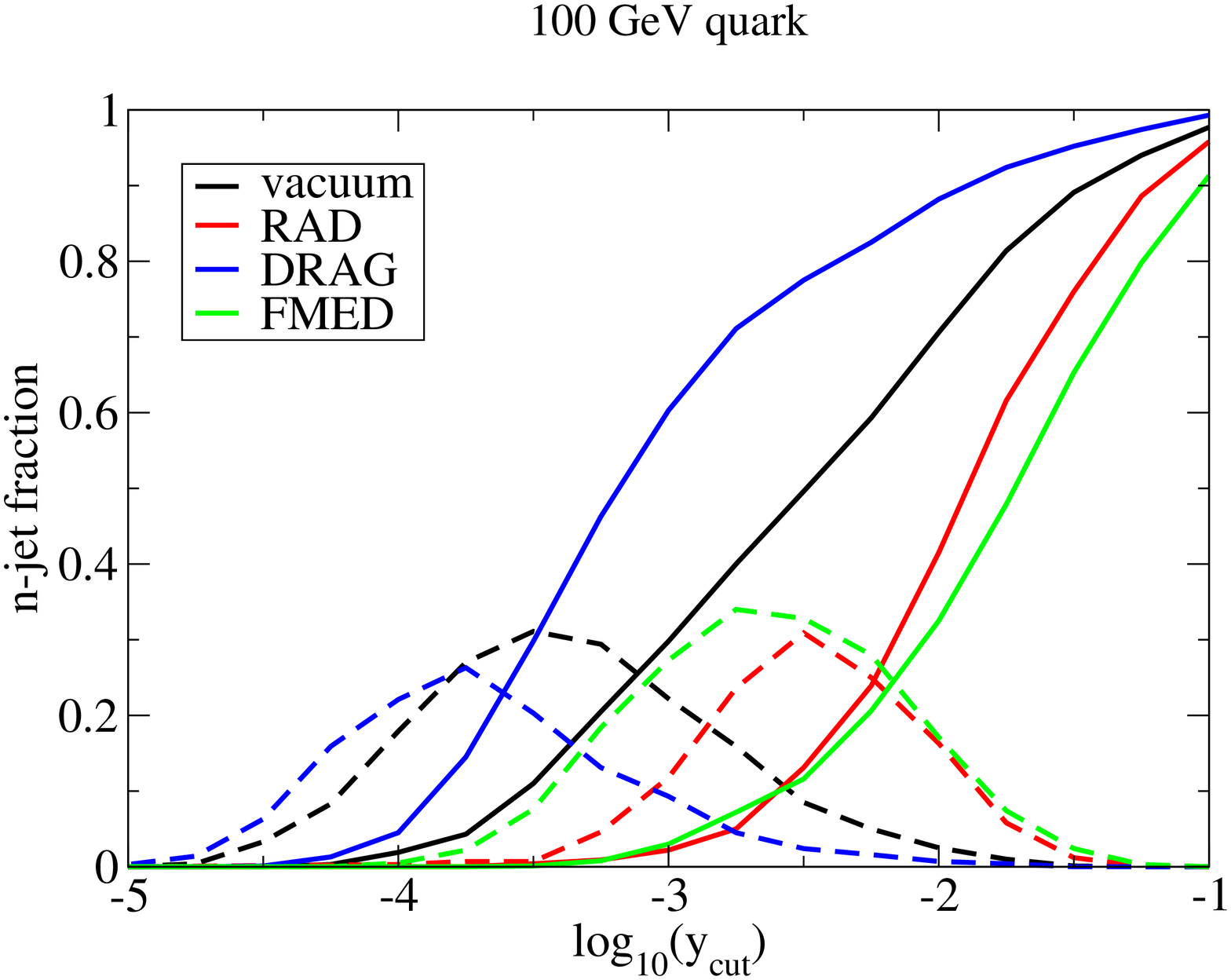}\includegraphics[width=8.0cm]{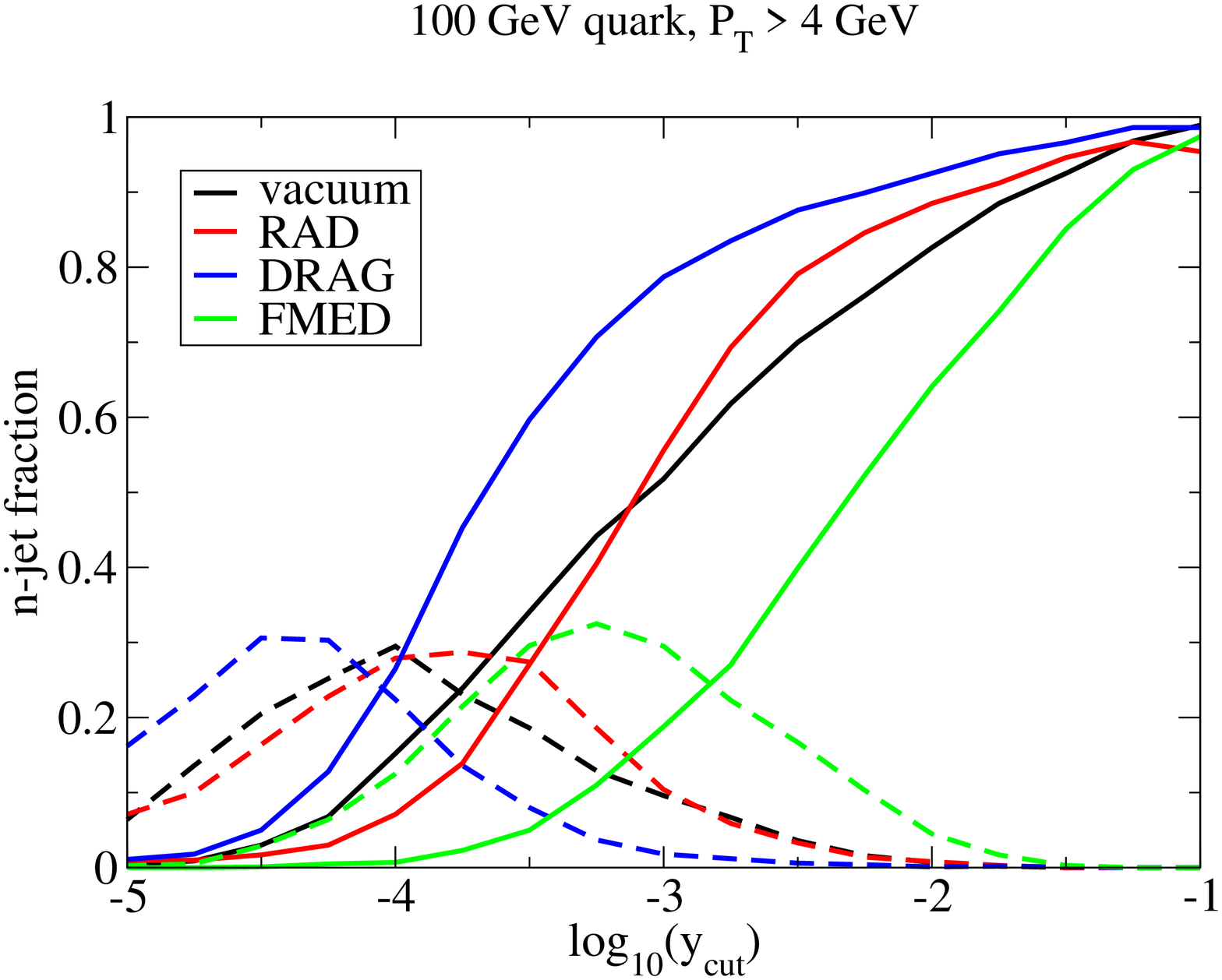}
\caption{$n$-jet fraction of dijets from a 100 GeV quark pair both in vacuum and medium modified as computed within YaJEM\cite{YaJEM-Jets}.}
\label{F-njet}
\end{center}
\end{figure}
As Fig.~\ref{F-njet} indicates, the modification of the branching pattern by the medium should be visible in a measurement of the $n$-jet fraction even above a $P_T$ cut.

Once the complications of jet measurement in a heavy-ion environment are sufficiently understood, there is some reason to believe that studying jets will eventually provide much more complete information about the dynamics of energy loss and redistribution than leading hadrons or correlations.

\subsection{Medium recoil from a hard probe}

Conceptually, if there is an interaction between a developing parton shower and a medium which can be described well by hydrodynamics, there must be a back-reaction of the medium to the perturbation in terms of shockwave excitation. Currently, such a picture is also indirectly supported by the success of the strong coupling scenarios for the parton-medium interaction in describing $R_{AA}(\phi)$ and $I_{AA}$ as well as by the absence of low $z$ multiplicity enhancement in $\gamma$-h correlations discussed above. This is a very tantalizing idea, as having a known localized perturbation in the system allows to measure yet more medium transport coefficients (such as the speed of sound) by observing the medium response carefully.

\begin{figure}[ht]
\begin{center}
\includegraphics[width=16.0cm]{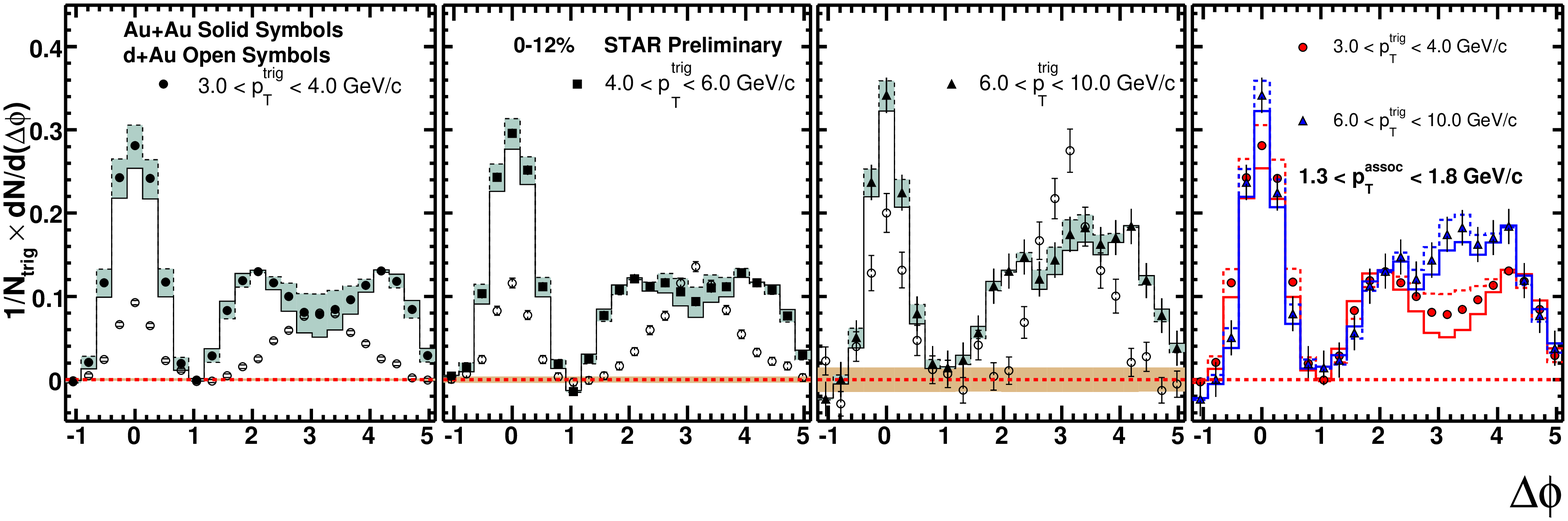}
\caption{Back-to-back correlation for different ranges of trigger momentum as measured by the STAR collaboration, shown for both d-Au and Au-Au data. At low trigger $P_T$ (left panel), the distortion of the correlation to a double-hump structure as compared to the d-Au Gaussian structure is clearly apparent.}
\label{F-dhump}
\end{center}
\end{figure}

It is currently an open question if such shockwaves have also been observed directly in correlation measurements with the associate hadrons observed at lower momenta. Fig.~\ref{F-dhump} shows such correlations.
Especially at low $P_T$, a double-hump structure is seen on the away side which would be at least consistent with the signal expected from a sonic shockwave. A number of proof-of-concept calculations of particular source terms inserted into ideal fluid dynamics have been performed so far (see e.g. \cite{Shock1,Shock2}), but more phenomenological models indicate that the proper averaging including the trigger surface bias and the coupling of the propagating shock to the medium flow field \cite{Mach1,Mach2,Mach3} as well as the detailed spatio-temporal structure of the source term \cite{Mach4} all have a critical influence.

Thus, while it appears that there is a possibility that shockwaves are seen in the data, the question is by no means settled and the theoretical understanding of the truly hard part of energy loss has not yet progressed to the level that one could attempt to reliably deduce a speed of sound from the observations. Whether shockwaves will be an issue for the LHC heavy ion program remains to be seen.

\section{Summary}

Despite knowing the Lagrangian, we do not really know much about QCD - away from the perturbative limit, even qualitative understanding of the implications of the Lagrangian is often absent. The aim of ultrarelativistic heavy-ion physics is to bridge this gap and to contribute to the understanding of thermodynamics and collective phenomena of QCD. As we have seen, for many observed phenomena it is not {\em a priori} clear in what degrees of freedom a model should be formulated, and only after experimental evidence it became clear that hot and dense QCD matter can be described as a fluid.

Hard probes serve in this context to provide information and constraints for models of hot QCD matter which can not be obtained by studying bulk matter alone. The observation that one can use hard processes as a standard-candle to image the medium via the final state interaction of outgoing partons is at the heart of the idea of tomography. This however requires careful and comprehensive modelling of all aspects of the dynamics of heavy-ion collisions.

As we have seen, the RHIC experiments have obtained already some measure of tomographic information, but in many cases this should be regarded as a proof of concept rather than a systematic investigation. Hard probes are the true domain of the LHC kinematic range, and precision high $P_T$ data from the LHC experiments is eagerly expected from the heavy-ion community.

\section*{Acknowledgements}

\end{document}